\documentstyle[11pt,adassconf]{article}  

\begin{document}   

%
%
%

\paperID{P2-13}

%
%

\title{PICsIM - the INTEGRAL/IBIS/PICsIT Observation Simulation Tool for Prototype Software Evaluation}

%
%
%

\author{J.\ B.\ Stephen and L.\ Foschini}
\affil{Istituto TeSRE/CNR, Via P. Gobetti 101, 40129 Bologna, Italy}

%
%

\contact{John B. Stephen}
\email{stephen@tesre.bo.cnr.it}

%
%
%

\paindex{Stephen, J.B.}
\aindex{Foschini, L.}     

%
%



\begin{abstract}          
The INTEGRAL satellite is an observatory-class gamma-ray telescope due for launch in early 2002. It comprises two main instruments, one optimised for imaging (IBIS) and the other for spectroscopy (SPI). The PICsIT telescope is the high energy (150 keV - 10 MeV) plane of the IBIS imager and consists of 8 individual modules of 512 detection elements. The modules are arranged in a 4 x 2 pattern, while the pixels are in a 16 x 32 array. This layout, which includes a dead area equivalent to one pixel width between each module, together with the event selection procedure, which (in standard mode) does not allow the identification of coincidences between separate modules, leads to a non-uniformity of the background which is significantly different for single-site events and for multiple energy deposits. Other sources of  background variations range from the separate low energy detector, situated immediately above the PICsIT plane, to the large mass of the SPI telescope at a short distance to one side. The algorithms for performing all the imaging and spectral deconvolution for PICsIT are currently being produced for delivery to the INTEGRAL Science Data Centre. In order to maximise the information which may be extracted from the PICsIT data, we have designed a prototyping environment which consists of a GUI to a highly structured and modular set of procedures which allows the easy simulation of observations from the data collection phase through to the final image production and analysis.
\end{abstract}

%
%

\section{Introduction}
The INTEGRAL high energy gamma-ray satellite consists of two primary instruments - a high resolution spectrometer (SPI) and a wide spectral and angular coverage imager (IBIS). IBIS achieves the goal of being able to create images of the gamma-ray sky over the energy range of $\approx$15 keV - 10 MeV with good angular resolution and large field of view by utilising two position sensitive detection planes, one solid state (ISGRI) optimised at lower energies and the other using scintillator technology (PICsIT) for high energy detection, in conjunction with a coded aperture mask. There are many possible sources of disturbance in the imaging process for the PICsIT detector, ranging from the presence of ISGRI between PICsIT and the coded aperture, through the modular form of the detection plane construction to the masses of the other instrumentation nearby.
\section{The PICsIM Environment}
The PICsIM simulation tool is written entirely using the IDL language (IDL 5.3, Research Systems Inc., Boulder, Colorado USA) and runs under both Windows NT and Linux operating systems. The tool itself provides only minimum functionality, mainly in the displaying of the images produced, however it is the interface which allows the developer straightforward and consistent control of the algorithms he is creating. This is performed by using a rigid system of data and algorithm storage which allows the operator to use the tool to identify the set of procedures available for use at any one time (e.g. for deconvolution of the image or definition of the accumulation parameters), and to select the required action. In this way it is also possible to bypass the data simulation step (which is itself a limited Monte-Carlo process which does not take into account all physical processes, and to use the results of a detailed Monte-Carlo simulation which is also being developed at the TeSRE institute (Malaguti et al. 2000) merely by placing the event files so produced into the event file directory and ensuring that a run-ID and associated statistics file is associated with them. Furthermore, although to date the various procedures used to define the imaging process are actually IDL source code, in principle they could also be pre-compiled C or FORTRAN programs called by the main IDL procedure. 

The entire PICsIM simulation, display and evaluation package is controlled through use of IDL widgets. The top level is the main control GUI which then activates one of the other second-level GUI's on user command. The various commands currently available are detailed below:

\begin{figure}
\epsscale{0.5}
\plotone{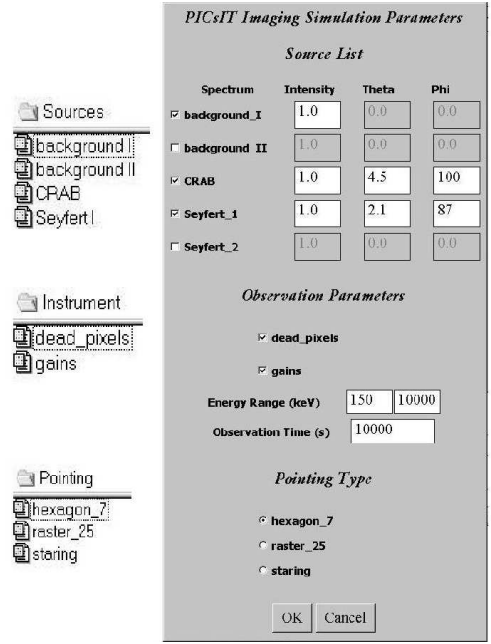}
\caption{The GUI to the 'SIMULATE' sub-menu}
\end{figure}

\begin{itemize}

\item{SIMULATE} This command activates the GUI which allows the simulation of an observation.
The simulation widget interface is shown in Figure 1. 
The source definition procedures (user defined) are located in the relevent directory and displayed, as are the instrument characteristic procedures which describe how the hardware is functioning and the observation strategy (e.g. hexagonal pointing, 'staring' etc.) Once a valid combination is selected a script file is created and the simulation automatically launched, creating a series of event files, one for each pointing, with a run-id assigned which is unique for every observation.

\item{ACCUMULATE} The event files form the input for this GUI which is shown in Figure 2. 
\begin{figure}
\epsscale{1.0}
\plotone{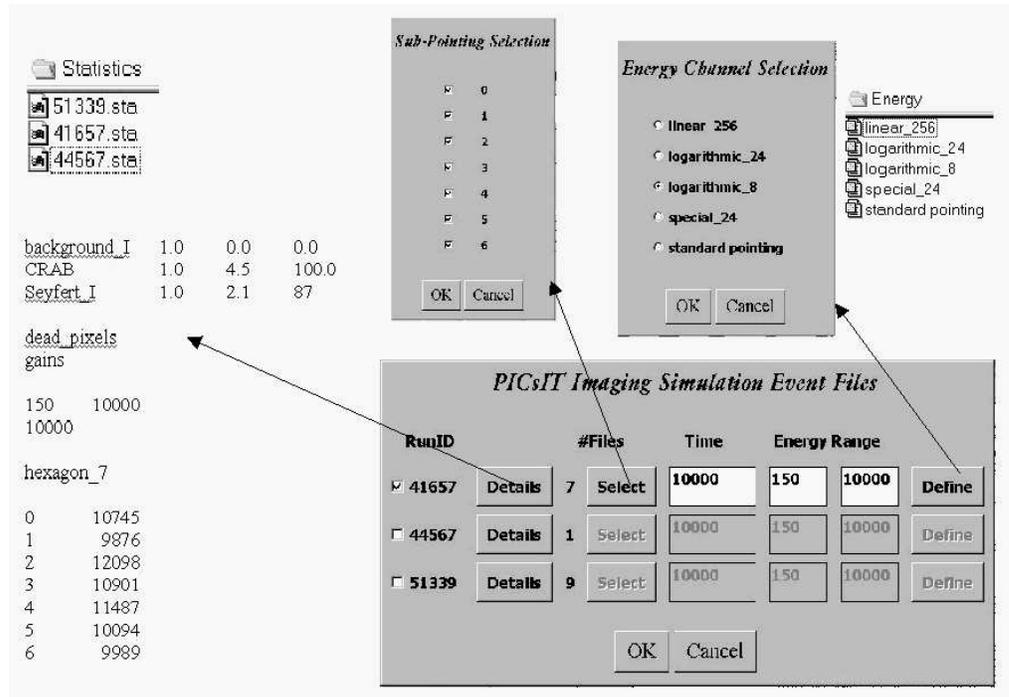}
\caption{The GUI to the 'ACCUMULATE' sub-menu}
\end{figure}
The user can select the files which he wants to accumulate by run-id (i.e. observation), and further detail may be imposed by selecting also by sub-pointing. The effective time may also be changed and the energy channels of the accumulation may be defined by means of selection of a user written procedure. Details of a particular observation may be obtained (these are held in the associated statistics file)

\item{DECONVOLVE} The deconvolution GUI allows the user to select between various methods of deconvolution (provided as IDL procedures in the appropriate directory) for the accumulated data files.

\item{CLEAN} This GUI screen shows all the options open (again in terms of IDL procedures) to the user for cleaning the resulting image, or filtering the accumulated data files before deconvolution.

\item{DISPLAY} This GUI provides a workspace within which the operator can display and perform limited analysis on any accumulated shadowgram, deconvolved and/or cleaned image or sum of images. Figure 3 shows how the operator can display the original detector image, the corrected shadowgram and the cleaned sky image of a point source.
\begin{figure}
\epsscale{1.0}
\plotone{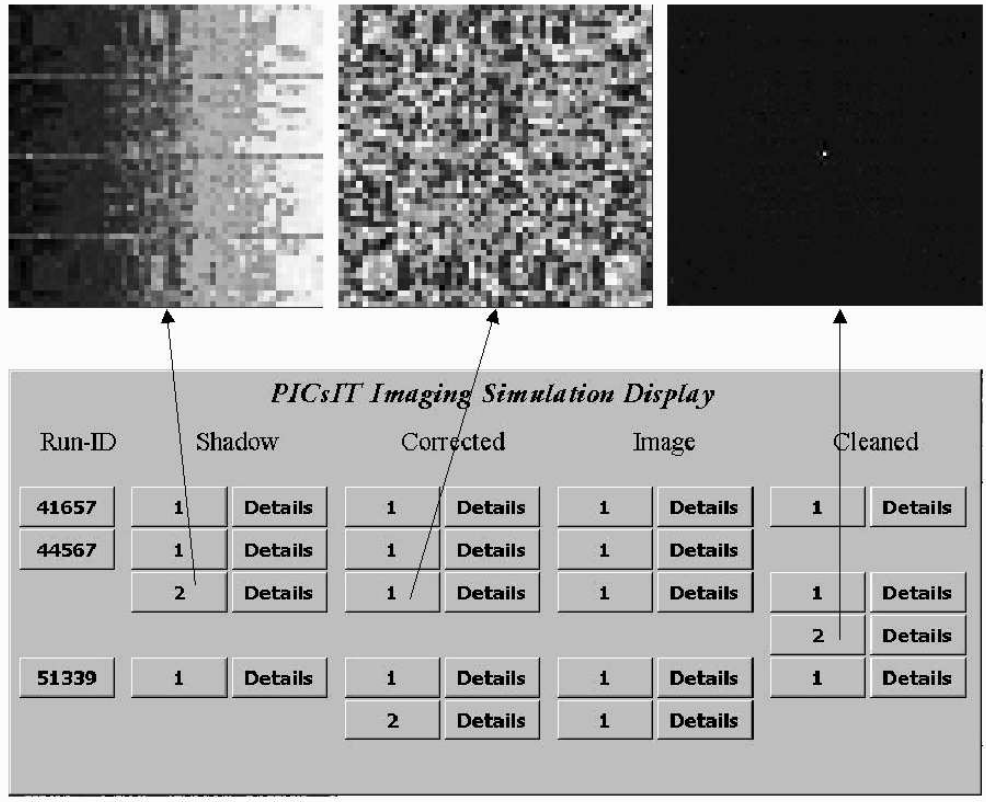}
\caption{The GUI to the 'DISPLAY' sub-menu}
\end{figure}
\end{itemize}

\section{Conclusions}
The PICsIM simulation environment is a useful tool within which the algorithms are being developed for use in the final data analysis data package at the ISDC. Its highly modular form and easy widget-driven interface allows a rapid evaluation and comparison of new procedures to be made.


\begin{references}
\reference Malaguti, G., Ciocca, C. and Di Cocco, G.  To be published in Proceedings of the 4th INTEGRAL workshop, Alicante, Spain, September 2000.
\end{references}
\end{document}